*Article*

# A Single-Frame and Multi-Frame Cascaded Image Super-Resolution Method

Jing Sun [1], Qiangqiang Yuan [2], Huanfeng Shen [1,*], Jie Li [2] and Liangpei Zhang [3]

1. School of Resource and Environmental Sciences, Wuhan University, Wuhan 430079, China; rainsunny@hotmail.com
2. School of Geodesy and Geomatics, Wuhan University, Wuhan 430079, China; qqyuan@sgg.whu.edu.cn (Q.Y.); aaronleecool@whu.edu.cn (J.L.)
3. The State Key Laboratory of Information Engineering in Surveying, Mapping and Remote Sensing, Wuhan University, Wuhan 430079, China; zlp62@whu.edu.cn
* Correspondence: shenhf@whu.edu.cn

**Abstract:** The objective of image super-resolution is to reconstruct a high-resolution (HR) image with the prior knowledge from one or several low-resolution (LR) images. However, in the real world, due to the limited complementary information, the performance of both single-frame and multi-frame super-resolution reconstruction degrades rapidly as the magnification increases. In this paper, we propose a novel two-step image super resolution method concatenating multi-frame super-resolution (MFSR) with single-frame super-resolution (SFSR), to progressively upsample images to the desired resolution. The proposed method consisting of an L0-norm constrained reconstruction scheme and an enhanced residual back-projection network, integrating the flexibility of the variational model-based method and the feature learning capacity of the deep learning-based method. To verify the effectiveness of the proposed algorithm, extensive experiments with both simulated and real world sequences were implemented. The experimental results show that the proposed method yields superior performance in both objective and perceptual quality measurements. The average PSNRs of the cascade model in set5 and set14 are 33.413 dB and 29.658 dB respectively, which are 0.76 dB and 0.621 dB more than the baseline method. In addition, the experiment indicates that this cascade model can be robustly applied to different SFSR and MFSR methods.

## 1. Introduction

High-resolution (HR) images with high perceptual quality are often required in applications such as video surveillance [1,2], face recognition [3], medical diagnosis [4], and remote sensing [5–7]. However, due to the different capabilities of sensors, the quality of captured images can vary greatly and fail to meet the requirements of subsequent applications. Super-resolution technology is an effective way to overcome the inherent resolution limitation of the current sensor imaging systems [8]. The objective of the super-resolution technique is to reconstruct an HR image from single or multiple LR observation frames captured at different perspectives of the same scene. In general, the observed LR image can be modeled as a degraded representation of the HR image, which are degraded by warp, blur, noise, and decimation [5]. According to the number of input LR images, the conventional super-resolution approaches can be roughly categorized into single-frame super-resolution (SFSR) [9–15] and multi-frame super-resolution (MFSR) [16–20].

Multi-frame super-resolution reconstruction aims to merge the complementary in-formation from different images to generate a higher spatial resolution image. The problem was first formulated by Tsai and Huang [16] in the frequency domain to improve the spatial resolution of Landsat Thematic Mapper (TM) images. Over the past few decades, the research work has been presented and studied in the spatial domain to improve multi-frame super-resolution techniques [17,18]. The SR problem is considered as an ill-posed inverse problem, as for each LR image, the space of its plausible corresponding HR images is huge and scales up quadratically with the magnification factor [21]. Owing to its effectiveness and flexibility, most research has focused on regularized frameworks, which impose some constraints on the solution space [22]. The maximum a posteriori estimation (MAP) method transforms the super-resolution reconstruction into an energy function optimization problem. Generally, the energy function consists of a data fidelity term that measures the model error between the degraded observations and the ideal image, and a regularization term that imposes some prior knowledge to constrain the model to achieve a robust solution. However, the priors of these methods are hand-crafted based on limited observations of specific image statistics, which may restore unsatisfactory results, as the real constraint often deviates from the predefined priors. On the one hand, the ill-posed nature is particularly evident for large magnification factors, which increases the problem of sub-pixel alignment and leads to the absence of texture details in the reconstructed images. On the other hand, it is difficult to obtain sufficient LR images with non-redundant information to recover the aliasing high-frequency components. Therefore, the performance of MFSR algorithms decreases rapidly with increasing magnification.

The mainstream algorithms of SFSR involve, e.g., reconstruction-based [9], example-based [23], sparse representation-based [24], regression-based [11], and deep learning-based approaches [13–15,25]. With the rapid development of deep learning, the convolutional neural network (CNN) dominated the research of SR due to its promising performance in terms of effectiveness and efficiency [26]. A pioneering work of SRCNN [12] applied a three-layer network to learn non-linear mapping relationships between the HR patches and the corresponding LR patches. Since then, considering the excellent learning capacity of convolutional neural networks (CNNs), deep learning-based methods have been developed in various ways by using new architectures or proper loss functions. The improved network [13] exploited residual learning (VDSR) [27] and recursive structure layers (DRCN) [28] to achieve an outstanding performance for SFSR. The residual dense network (RDN) [14] innovatively combined residual learning and dense connection to fully utilize both the shallow features and deep features together with over 100 layers. Recently, the network of channel attention (RCAN) [29] and second-order channel attention (SAN) [15] were introduced to exploit feature correlation for superiority performance. These end-to-end networks compute a series of feature maps from the LR image, culminating with one or more upsampling layers to construct the HR image. Therefore, it is convenient in that it automatically learns good features from massive quantities of data without much expertise and manual feature learning. Nevertheless, many deep learning approaches hypothesize that the training and test dataset are drawn from the same feature space with the similar distribution. Hence, the SR performance is heavily bound to the consistency between testing data and training data [8]. Meanwhile, learning-based methods directly generate high-resolution details according to the learned mapping functions and low-resolution input, and some unexpected artifacts may be produced in the reconstructed results, especially for large magnification factors. Furthermore, the difficulty in estimating missing high-frequency details increases with the scale factor due to the increment in the ambiguities between LR and HR.

Briefly speaking, the SR performance at a large scale factor remains a challenging problem for both the MFSR and SFSR approaches. On the one hand, model-based MFSR algorithms encounter difficulty in recovering missing high-frequency details with the limited complementary information. On the other hand, at large upsampling scales, since insufficient information is available to recover such high-frequency components, deep learning-based SFSR methods may "hallucinate" the fine detail structure. In particular, the hallucination can be very problematic in some critical applications. To deal with this challenge, some researchers [30,31] have proposed exploiting the complementary advantages of external and internal information to improve SR performance and perceptual visual quality. However, most deep learning-based video and multi-frame super-resolution methods cannot fully exploit the temporal and spatial correlations among multiple images. Their fusion modules do not adapt well to image sequences with weak temporal correlations [32]. These methods cannot satisfy our everyday requirements, because of the limited information involved in the reconstruction model.

To our knowledge, the MFSR and SFSR methods extract missing details from different sources. SFSR extracts various feature maps representing the details of a target image. MFSR provides multiple sets of feature maps from other images. The model-based MFSR methods and the deep learning-based SFSR procedures are complementary, to a large extent [33]. Combining the feature learning capacity of SFSR with the information fusion brought by MFSR, a few pieces of research proposed a combination of single-frame and multi-frame SR such as [34,35]. In [34] the input LR images are first magnified and recovered by a conventional MFSR method with a 4× scaling factor; then, an SFSR network is applied to the previous recovered result for artifacts removal without magnification. The authors of [35] carried out the process in the inverse order to [34], where they input LR images separately through the SFSR network, and then a conventional MFSR was applied on the resulting image. In contrast, the SFSR network in the former framework is only used as a filter to fine-tune the output of the MFSR method, while the SFSR network is used to initialize the input of the MFSR method in the latter research. Compared with traditional methods, the cascade model can simultaneously capitalize on both inter-frame aliasing information and external learned feature information, which notably improves the utilization of multiple images and external example data.

In this paper, we propose a novel two-step super-resolution reconstruction method concatenating the L0-norm constrained reconstruction with an enhanced residual back-projection network. Such a cascade model property induces considerable advantages for image SR, which integrates the flexibility of model-based method and the feature learning capacity of learning-based method. Specifically, the L0-norm constrained reconstruction method takes multiple images as input to obtain an initial high-resolution image, and then an enhanced residual back-projection network is further applied to the initial image for recovering a more accurate result. The proposed cascade model leverages the information learned from multiple low-resolution inputs and neural networks, outperforming the existing baseline SR methods in the cascade model in both objective and perceptual quality measurements.

The rest of this paper is organized as follows: Section 2 introduces the variational model-based MFSR algorithm and the deep learning-based SFSR algorithm that are concatenated in the cascade model. We present the detailed experimental results for this multi/single-frame super-resolution cascade model in Section 3, followed with a discussion of the strategy for cascade model in Section 4. Finally, our conclusions are drawn in Section 5.

## 2. The Cascade Model for Image Super-Resolution

Most methods reconstruct HR images in one upsampling step, which increases the difficulty of reconstructing at large scaling factors. A Laplacian pyramid framework (LapSRN) [36] is proposed to progressively reconstruct multiple images with different scales in one feed-forward. However, this network relies only on the limited features available in the LR space with a stack of single upsampling networks. Because of the insufficient information available to restore such high frequencies, it is unrealistic to generate sharp HR images with fine detail at large scale factors.

The cascade model of MFSR and SFSR is proposed to obtain high-performance results for image super-resolution at large scaling factors. There are four structures for performing SR using MFSR, SFSR, or combinations of them when the upscaling factor is a divisible integer such as 4, as shown in Figure 1. To the best of our knowledge, the question of how to best combine SFSR and MFSR has not been answered theoretically. Since the actual degradation is more complex and varying, the learning-based SFSR cannot fully simulate the image degradation process, which may cause incorrect results in actual reconstruction. In order to reduce the error transmission, we suggest using the multi-frame first and then single-frame cascade method for super-resolution (MFSF-SR), while the opposite method by applying SFSR first and MFSR after (SFMF-SR) is analyzed in detail in the subsequent discussion section.

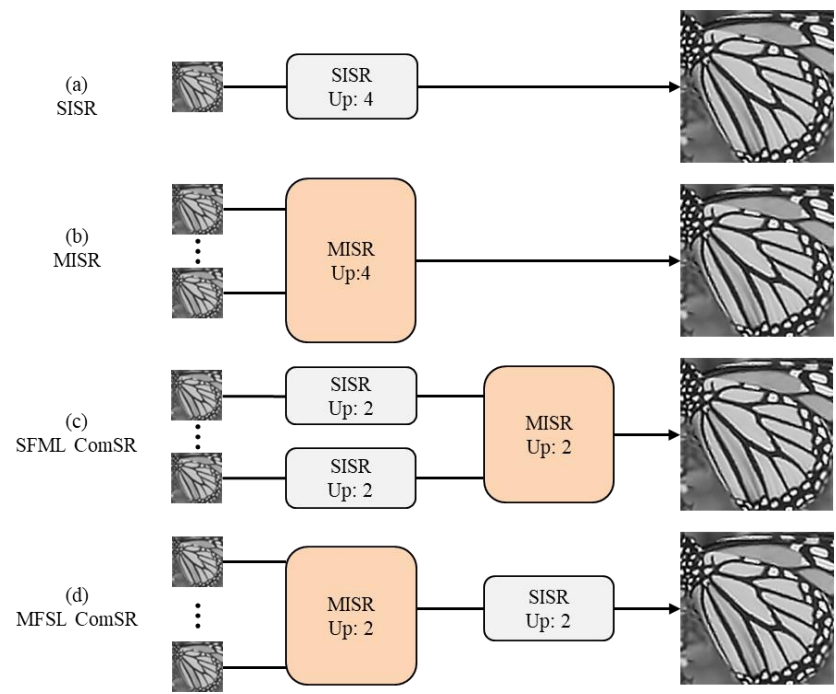

**Figure 1.** Various structures of the super-resolution cascade model.

The proposed cascade method consists of two main parts: the variational model-based MFSR and the deep learning-based SFSR. We aim to concatenate the MFSR method with the SFSR method to progressively upsample images to the desired resolution. Regarding the choice of the MFSR and SFSR methods, we employ the MFSR approach via L0-norm regularized intensity and gradient combined prior (L0RIG) and the SFSR approach using enhanced residual back-projection networks (ERBPN), respectively, which are introduced in the following subsection.

*2.1. Multi-Frame Super-Resolution via the L0-Norm Regularized Intensity and Gradient Combined Prior*

In image super-resolution reconstruction, as a typical inverse problem, SR is highly coupled with the degradation model. Generally speaking, the HR image is inevitably corrupted by many factors in the acquisition process, including warping, blurring, sub-sampling operators, and additive noise [5]. It allows for us to reconstruct an output image above the Nyquist Limit of the original imaging device. Super-resolution turns out to be an inherently ill-posed inverse problem because the information contained in the observed LR images is not sufficient to solve the HR image. Therefore, it is necessary to impose a specific regularization in order to obtain a stable solution. The model-based methods incorporate prior constraints to estimate the desired HR image by minimizing an objective function of the posterior probability.

We denote the ideal HR image required to be reconstructed as $z \in R^{M_s \times N_s}$, the observed LR images as $\{g_k\}_{k=1}^{s^2} \in R^{M \times N}$, the downsampling matrix as $D \in R^{MN \times MNs^2}$, the motion matrix as $\{M_k\}_{k=1}^{s^2} \in R^{MNs^2 \times MNs^2}$, and $B \in R^{MNs^2 \times MNs^2}$ as the blur matrix including the sensor blur, optical blur, and atmospheric turbulence, where we assume that the blur of multiple images obtained under the same scene is consistent. The additive noise of the image observation model is usually assumed to be white Gaussian noise. Thus, the size of the LR image $g_k$ is $M \times N$, the scaling factor is $s$, and the size of the HR image is $M_s \times N_s$. By changing the number of LR images, they can be applied to the MFSR or SFSR tasks. The MAP-based solution model for the super-resolution problem can be represented by a generalized minimization cost function as follows:

$$\hat{z} = \arg\min_z \left\{ \sum_k^K \|g_k - DBM_k z\|_2^2 + \lambda U(z) \right\} \quad (1)$$

The first term of the cost function is the data fidelity term, which measures the reconstruction error to ensure that pixels in the reconstructed HR image are close to real values; the second term $U(z)$ is the regularization term associated with the general prior information about the desirable HR image to obtain a robust solution; and $\lambda$ is the regularization parameter, which provides a tradeoff between the data fidelity term and the regularization term.

In the image processing field, Gaussian-type noise is the most commonly assumed because the noise generated in image acquisition usually satisfies a Gaussian distribution [22]. We assume the noise to be additive white Gaussian noise, so the fidelity term can be characterized by the L2-norm. For the regularization term, Laplacian [37], total variation (TV) [20] and Huber–Markov random field (HMRF) [38] regularization are first considered, due to their simplicity and efficiency. Based on the advantages of the TV regularization, a combined image prior based on intensity and gradient is proposed for natural images [39], which describes the two-tone distribution characteristics of the gradient statistics. This expression is written as follows:

$$U(z) = \|z\|_0 + \|\nabla z\|_0 \quad (2)$$

where $\nabla$ is the gradient operator. As the intensity prior is based on independent pixels instead of the disparities of neighboring pixels, it introduces significant noise and artifacts in the image restoration. In contrast, the gradient prior is based on the disparities of neighboring pixels, and thus enforces smooth results with fewer artifacts. Prior knowledge for constraining the intensity and gradient can sufficiently exploit the statistical properties of natural images. To effectively preserve the detailed texture information and enhance the reconstructed image quality, the intensity and gradient combined prior is employed in the super-resolution reconstruction [40]. We propose an MFSR algorithm via an L0-norm regularized intensity and gradient combined prior (L0RIG) to integrate into the cascade model.

Typically, geometric registrations and the blur can be estimated from the input data and used with the generative model to reconstruct the super-resolution image. The super-resolution becomes very limited without a good estimation of the blur and motion between the LR sequences. In this work, we compute the warping matrix $M$ and blur matrix $B$ with the optical flow approach [41] and the blind blur kernel estimation method [39], respectively. In order to simplify Equation (1), $DBM_k$ can be regarded as a system matrix $W_k$. By substituting Equation (2) into Equation (1), the following minimization function for solving the MFSR model can be obtained:

$$\hat{z} = \arg\min_z \left\{ \sum_k^K \|g_k - W_k z\|_2^2 + \lambda (\|z\|_0 + \|\nabla z\|_0) \right\} \quad (3)$$

Due to the L0 regularization term in Equation (3), it is difficult to solve the super-resolution model since it is a nonconvex function. As known, variable splitting and alternate iterative optimization algorithms are typically used for optimizing the solutions of the variational model. Based on the variable splitting L0 minimization approach, we adopt the alternating direction method of multipliers (ADMM) algorithm [42] to solve the model. We introduce the auxiliary variables $u$ and $v$, representing $z$ and $\nabla z$, respectively, to move a few terms out of the non-differentiable L0 norm expression. The objective function can be rewritten as follows:

$$\hat{z} = \arg\min_z \left\{ \sum_k^K \|g_k - W_k z\|_2^2 + \lambda (\|u\|_0 + \|v\|_0) \right\} \quad s.t. \ u = z, v = \nabla z \quad (4)$$

By transforming Equation (4) to generate an unconstrained problem with the augmented Lagrangian algorithm, it can be rewritten:

$$\hat{z} = \arg\min_z \left\{ \sum_k^K \|g_k - W_k z\|_2^2 + \frac{\beta}{2}\|z - u\|_2^2 + \frac{\mu}{2}\|\nabla z - v\|_2^2 + \lambda(\|u\|_0 + \|v\|_0) \right\} \quad (5)$$

where $\beta$ and $\mu$ are penalty parameters, and are set to be 0.001 initially, that times 0.9 after each iteration to accelerate the convergence. Equation (5) can be efficiently solved through alternately minimizing $z$, $u$, and $v$ independently, by fixing the other variables. The flowchart of the MFSR via L0-norm regularized intensity and gradient combined prior (L0RIG) algorithm is illustrated in Figure 2.

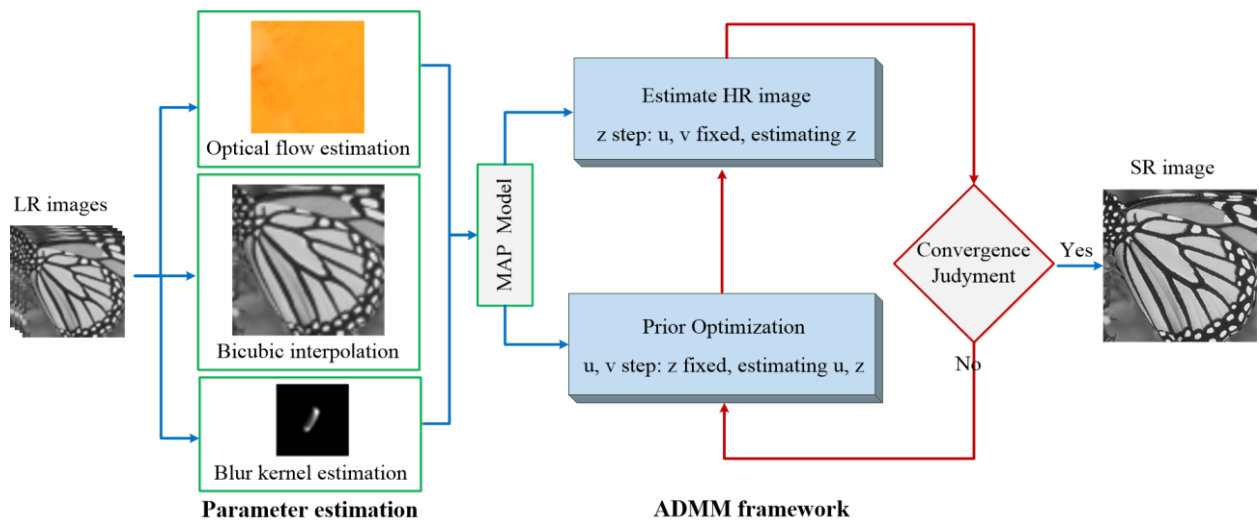

**Figure 2.** Flowchart of L0RIG.

## 2.2. Single-Frame Super-Resolution using Enhanced Residual Back-Projection Network

Inspired by the idea of iterative back-projection framework, Haris et al. [43] proposed deep back-projection network (DBPN) to iteratively use error feedbacks from the multiple up- and downscaling steps, which achieves the state-of-the-art SR performance with large scale factors. Since the iterative up/downsampling framework has the advantage of capturing the deep relationships between LR and corresponding HR images, it has become a promising framework in the field of SFSR [44]. Figure 3 illustrates the schematic pipeline of the proposed enhanced residual back-projection network (ERBPN), which is designed on the basic architecture of the original DBPN [43]. The architecture of ERBPN consists of three parts, namely, initial feature extract module, projection unit, and SR reconstruction module, as described below. Some modifications were made for the projection unit: (1) the down-projection unit was replaced with the downsampling unit; (2) the concatenation operation was replaced with a sequential feature fusion (SFF) operation. In the following, the major improvements are further explained.

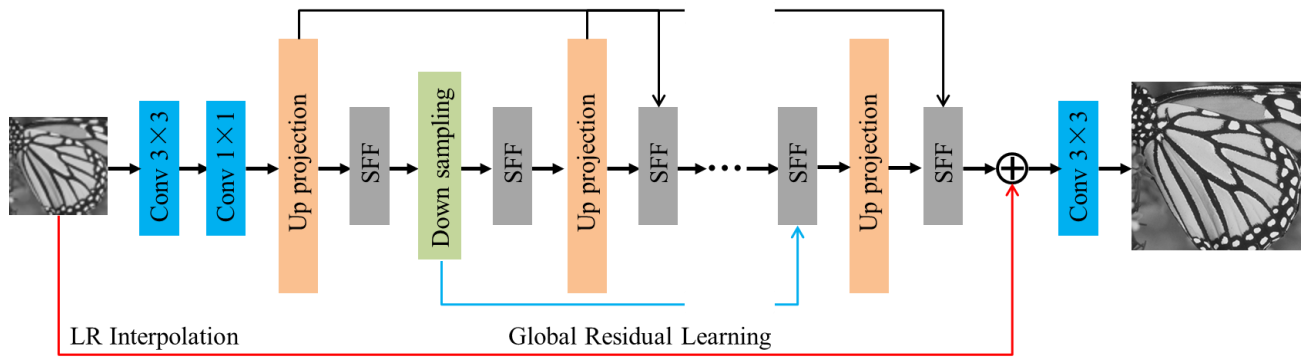

**Figure 3.** Architecture of ERBPN.

The first part extracts the shallow feature $L_0$ from the input LR image $I_{LR}$ and can be formulated by $L_0 = f_{init}(I_{LR})$, where $f_{init}$ denotes a convolution operation with $Conv(3, n_l, n_f)$ and $n_l$, $n_f$ are the number of input LR image channel and the feature maps, respectively. Then, a 1 × 1 convolution layer is used as feature pooling and dimension reduction before entering the projection unit.

Then, the initial feature extraction is followed by a sequence of projection units, alternating between construction of the LR and HR feature maps $L_t$, $H_t$. The projection units in our proposed framework include the up-projection unit and the downsampling unit. Iterative error feedback mechanism is proposed by iteratively estimating and applying a correction to the current estimation of the LR and HR feature maps. Here, the projection errors are used to characterize or constraint the features in early layers. The up-projection unit is utilized to map the LR feature maps to the HR feature maps, which is shown in Figure 4a. However, it is intuitive that obtaining LR feature maps from HR feature maps is simple and does not require projection unit based on iterative error feedback mechanism. Therefore, we simplify the back-projection network with a downsampling unit for faster computation, which has a very simple structure with a convolution layer as is shown in Figure 4b. Note that each input feature map is concatenated and fused through the sequential feature fusion (SFF) operation before entering the projection unit.

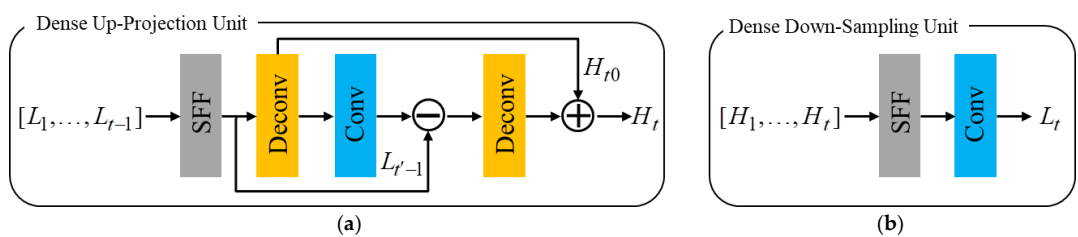

**Figure 4.** Architecture of the dense up-projection unit (**a**) and downsampling unit (**b**).

The up-projection and downsampling unit are densely connected to alleviate the vanishing gradient problem, produce improved feature, and encourage feature reuse [14]. The input for each unit is the concatenation of the outputs from all previous units to generate the feature maps effectively. Generally speaking, the feature maps generated by different projection units have different types of HR and LR components with different impacts on the quality of the results. Therefore, it is necessary to discriminate these feature maps with a feature fusion module [45]. In our framework, the sequential feature fusion operation (SFF) is employed to deal with the feature maps discriminatorily, integrating these feature maps in a sequential manner. Figure 5 shows the illustration of the SFF. Suppose that $m^t$ represents the $t$th input LR/HR feature map, $y^t$ denotes the output of the $t$th convolutional layer. Next, we obtain the following equation:

$$y^t = f([m^t; y^{t-1}]) \quad (6)$$

where $t = 1, 2, \ldots, n$, $y^0 = 0$. $n$ denotes the number of projection units, $[\cdot;\cdot]$ represents the concatenation operation, and $f$ denotes a convolution operation with 3 × 3 convolutional layer. It is worth pointing out that the SFF has discriminative ability because the feature maps generated by different projection units are processed at different depths of the network. Different from other networks, our reconstruction directly exploits different types of LR-to-HR features without propagating through up-projection layers.

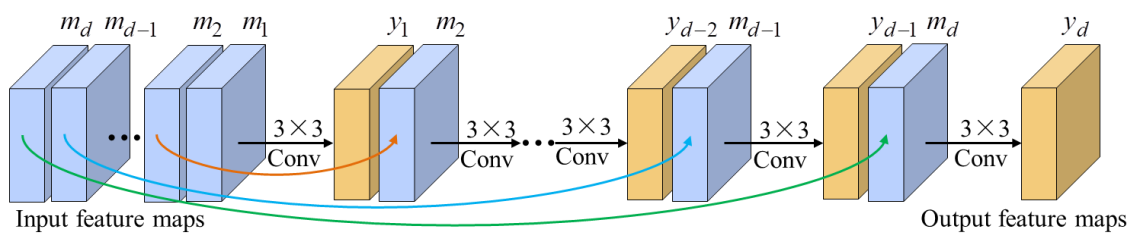

**Figure 5.** Architecture of the SFF.

Finally, we employ a global residual back-projection block structure. Residual learning helps the network converge faster and makes it easier for the network to generate only the difference between the HR and interpolated LR images [29], which can address the performance degradation problem caused by the details loss after so many layers in deep networks. In our ERBPN framework, the LR image is taken as the input to reduce the computation time. At the last stage, all HR feature maps from the up-projection step are deeply concatenated and fused with the SFF, then added to the interpolated LR image to generate the final super-solved image.

The last convolution layer is used for image reconstruction with filter size of 3 × 3. The network takes the reconstructed results, denoted as $z'$, as the output. Loss functions help us estimate the difference between the recovered SR images and the corresponding ground-truth HR images. MSE loss between the ground-truth HR image and the reconstructed HR image is used as the objective function, which can be written as follows:

$$Loss = \frac{1}{N} \sum_{i=1}^{N} \|z_i - z'_i\|_2^2 \quad (7)$$

where $N$ is the number of the training images.

*2.3. Summary of the Proposed Cascade Model for Super-Resolution*

In our work, the two-step super-resolution reconstruction method cascades the model-based MFSR and the deep learning-based SFSR method abovementioned. The MFSR with L0-norm regularized intensity and gradient combination prior (L0RIG) and the SFSR via enhanced residual back projection network (ERBPN) are employed to reconstruct a more accurate result. Specifically, first, we take 16 low-resolution images as the input of the L0RIG method to reconstruct one intermediate super-resolved image denoted as $z^l$, whose dimensions are 2× larger than the input LR images. Then, the intermediate super-resolved image $z^l$ is fed into the ERBPN framework to obtain a high-resolution result $z^{l+}$ with better quality. The high-resolution result $z^{l+}$ are 2× larger than $z^l$, hence 4× larger than the input LR images. Even though we exemplify our super resolution reconstruction method using 4× scaling factor, it can be directly extended to other SR scaling factors. The schematic diagram for the proposed cascade method is illustrated in Figure 6.

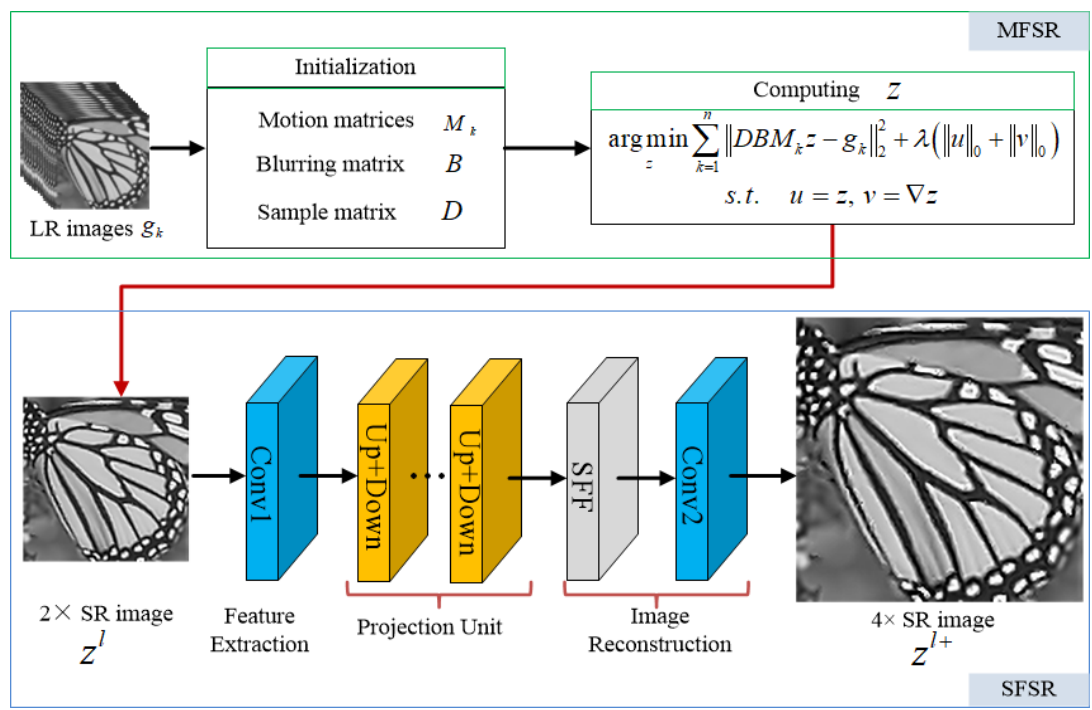

**Figure 6.** Flowchart of the super-resolution cascade model.

## 3. Experiments

To validly confirm the effectiveness of the proposed cascade model of MFSF-SR, this section presents the experimental results on both synthesized and real images. We combine the multi-frame-based L0RIG method with the single-frame-based ERBPN method, to up sampling images progressively at the 4× scale factor. The proposed cascade method applies ERBPN directly on the output of L0RIG in a sequential manner, where the L0RIG method reconstructs the LR images first, and then the resulting image is independently enhanced using the ERBPN method to obtain a higher-quality output. At the same time, the two baseline super-resolution reconstruct methods of L0RIG and ERBPN are also implemented to compare with the cascade method. In the simulation experiments, the effect of the proposed method under different noise levels is further investigated to verify the robustness to noise. The detailed steps are presented in the following sections.

*3.1. Data and Training Details*

The five grayscale HR images shown in Figure 7 were selected as the test images in the simulation experiments. For each image from these test sets, we generated a set of N =

16 images with different subpixel shifts applied before further degradation. Synthetic sequences of 16 LR images were generated by applying isotropic Gaussian blur to the sequential subpixel shifts HR image, then downsampling the row and column of the image by a factor of 4.

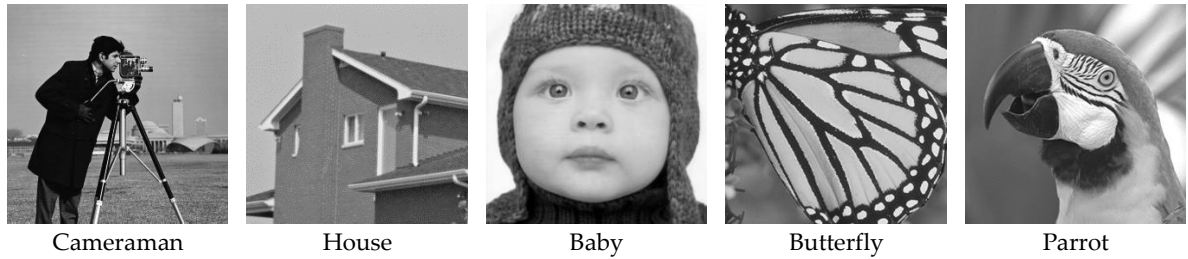

**Figure 7.** The five test images for image super-resolution.

In the reconstruction stage of the L0RIG, the central frame of LR sequence is chosen as our reference frame and the initial HR image is obtained by bicubic interpolation method. The regularization parameter $\lambda$ is determined empirically based on numerous experiments to produce the best performance. Since minimizing the objective function by preconditioned conjugate gradient method usually converges within 30 iterations, the maximum iteration number is set to TS = 30.

In the ERBPN, the filter size in the up-projection unit varies with respect to the scaling factor. For the 2× enlargement, we used a 6 × 6 convolutional layer with two striding and two padding. The 4× enlargement then used an 8 × 8 convolutional layer with four striding and two padding. In the training phase, we augmented the training data from the DIV2K dataset [46] by randomly employing 90°, 180°, and 270° rotation and horizontal and vertical flipping [44]. In each mini-batch, 128 degraded LR images with a patch size of 64 × 64 were provided as inputs for the model, and the corresponding HR image served as the ground truth for calculating the loss. The models were optimized using the ADAM optimizer [47] with $\beta 1 = 0.9$, $\beta 2 = 0.999$, and $\varepsilon = 10^{-8}$. The initial learning rate was set to $10^{-4}$ and then decreased by half every 100 epochs. A total of 1000 epochs were used for training the models since more epochs did not bring further improvements. All experiments were implemented using Caffe framework version 1.0.0-rc3 and MATLAB R2022a on an Nvidia RTX GPU, Santa Clara, CA, USA.

Image enhancement or visual quality improvement can be subjective because the perception of better image quality can vary from person to person. For this reason, it is necessary to establish quantitative measures for the comparison of image enhancement algorithms. To assess the image quality of the super-resolution reconstructed results, two classical evaluation criteria—the peak signal-to-noise ratio (PSNR/dB) and the structural similarity index measure (SSIM)—were chosen to measure the performance of the different super-resolution methods [48]. The higher the quantitative measure, the better the quality of the reconstructed image.

*3.2. Experiments on Synthetic Data*

L0RIG and ERBPN are the baseline methods of the proposed MFSF-SR, which only reconstruct by upsampling one step instead of step-by-step reconstruction under an upscaling factor of 4, for comparison with the cascade model. For a fair comparison, we run SFSR method for all 16 simulated LR images and compute the mean metric from the reconstruction outcomes—this way, the method is fed with the same data as those for MFSR. Additionally, a bicubic interpolation of the LR reference frame is also constructed for comparison.

Table 1 shows the quantitative performance comparison in terms of PSNR and SSIM for the five simulated images presented in Figure 7 with the different methods. For the sake of comparison, the two types of L0RIG and ERBPN algorithm directly reconstructed on 4× enlargement. The output of the cascade model is a super-resolved central frame with four times the size of the original LR images.

**Table 1.** Quantitative results (PSNR(dB) and SSIM) of the simulation experiments for 4× SR. The bold portion indicates the best performance.

| Data | Metric | Bicubic | L0RIG | ERBPN | MFSF-SR |
|---|---|---|---|---|---|
| Cameraman | PSNR | 24.120 | 26.004 | 26.787 | **27.642** |
|  | SSIM | 0.756 | 0.823 | 0.832 | **0.866** |
| House | PSNR | 27.228 | 31.572 | 32.549 | **33.391** |
|  | SSIM | 0.792 | 0.868 | 0.881 | **0.896** |
| Baby | PSNR | 31.770 | 33.653 | 34.352 | **34.744** |
|  | SSIM | 0.856 | 0.898 | 0.915 | **0.922** |
| Butterfly | PSNR | 22.099 | 25.073 | 26.006 | **26.863** |
|  | SSIM | 0.738 | 0.866 | 0.874 | **0.884** |
| Parrot | PSNR | 25.724 | 28.740 | 29.905 | **30.636** |
|  | SSIM | 0.874 | 0.916 | 0.935 | **0.941** |

For the sake of comparison, we analyzed the simulated experimental results from both subjective and objective perspectives. Quantitatively, as displayed in Table 1, the proposed cascade model yields the best scores in the evaluation metrics among all the compared methods. In the experiment with the butterfly image, the PSNR values are 25.073 dB for L0RIG, 26.006 dB for ERBPN, 26.863 dB for MFSF-SR. These quantitative results confirm the effectiveness of the MFSF-SR cascade model. From a subjective perspective, the red rectangles show zoomed regions of the restored images, to compare the qualitative performance of the different methods. L0RIG shows the preferable performance, but some edge is oversmoothed. ERBPN can produce good contrast through the up- and down-projection unit, but there are some unnatural artifacts around the slight edge. The result of the proposed MFSF-SR method contains more details and fewer blurred contours than L0RIG and ERBPN.

Furthermore, in the experiment with the parrot image, the PSNR value for the proposed MFSF-SR is 30.636 dB, which is 1.896 dB and 0.731 dB better than L0RIG and ERBPN, respectively. As displayed in Figure 8, images reconstructed with the MFSF-SR cascade model are able to preserve the HR components which contain more details, with rare additional artifacts. As a simple comparison, in the bottom line of Figure 8, the enlarged image in the result of L0RIG shows the misinterpreted area of the diagonal stripe due to the ringing artifact effect. It shows that the MFSF-SR can preserve the low-frequency content, and reliably restore the high-frequency details with the combination of the inter-frame information and external learning prior. From both the qualitative and quantitative analyses, most of the results show that the MFSF-SR with a two-step reconstruction creates more high-frequency information than the baseline methods at large magnification factors.

To further assess the robustness of the proposed method with regard to different noise levels, the Zebra image from the BSD68 dataset [49] was also selected as a synthesized test image with warping, blurring, downsampling, and different noise levels of additive white Gaussian noise (AWGN) added. For the color image sequence of the synthesized zebra image, we first convert the color input to YCbCr space, and then reconstructed the luminance component with the super-resolution algorithm.

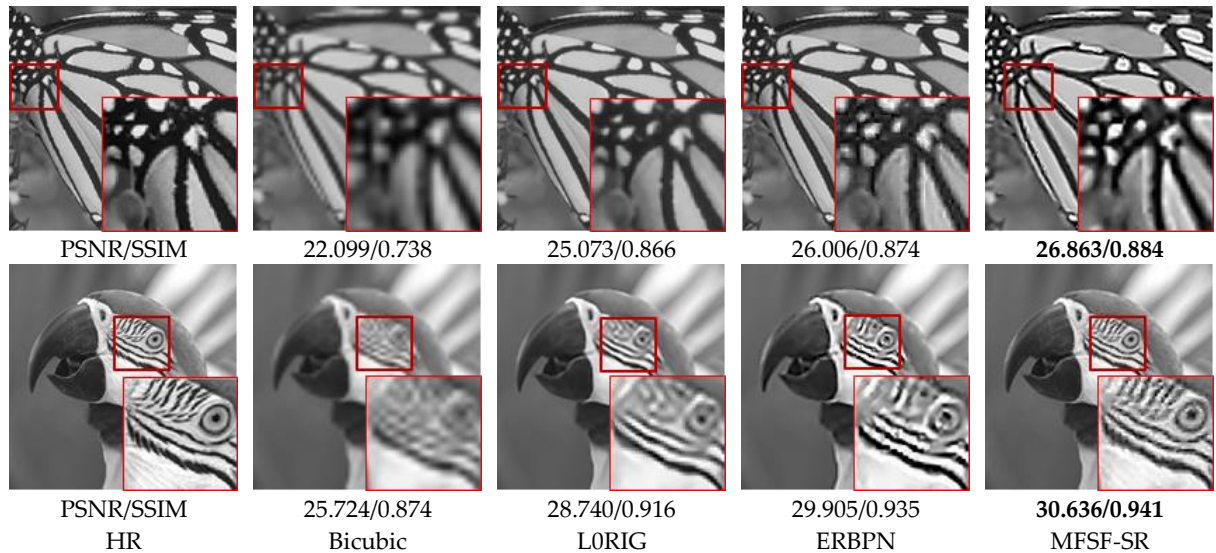

**Figure 8.** Qualitative comparison of the butterfly and parrot images on 4× SR.

To further compare the performance of the proposed method, a simulation experiment with the zebra image was implemented under different noise levels. The quantitative reconstruction results of the different methods with the color zebra image are shown in Table 2, where the proposed MFSF-SR method achieves very pleasing PSNR and SSIM results at all the noise levels. Figure 9 shows the quantitative performance comparison in terms of PSNR and SSIM for the zebra images under different noise levels. To be specific, in the experiment with a noise variance of 0.005, the proposed method outperforms all the compared methods with a result of 29.22 dB, which is 0.907 dB and 1.325 dB better than L0RIG and ERBPN, respectively. Furthermore, it can be observed that the performance advantage is more obvious for the high noise levels, and the proposed method turns out to be effectively adapted to different noise characteristics.

**Table 2.** Quantitative results of the simulation experiment with different noise levels for 4× SR. The bold portion indicates the best performance.

| Noise variance | Metric | Bicubic | L0RIG | ERBPN | MFSF-SR |
|---|---|---|---|---|---|
| 0.001 | PSNR | 19.698 | 22.538 | 22.518 | **23.206** |
| | SSIM | 0.783 | 0.901 | 0.899 | **0.917** |
| 0.002 | PSNR | 19.681 | 22.151 | 22.036 | **22.703** |
| | SSIM | 0.782 | 0.892 | 0.889 | **0.906** |
| 0.003 | PSNR | 19.666 | 21.825 | 21.673 | **22.341** |
| | SSIM | 0.781 | 0.884 | 0.881 | **0.896** |
| 0.004 | PSNR | 19.651 | 21.549 | 21.379 | **22.002** |
| | SSIM | 0.779 | 0.877 | 0.873 | **0.887** |
| 0.005 | PSNR | 19.638 | 21.313 | 21.095 | **21.822** |
| | SSIM | 0.778 | 0.872 | 0.866 | **0.881** |

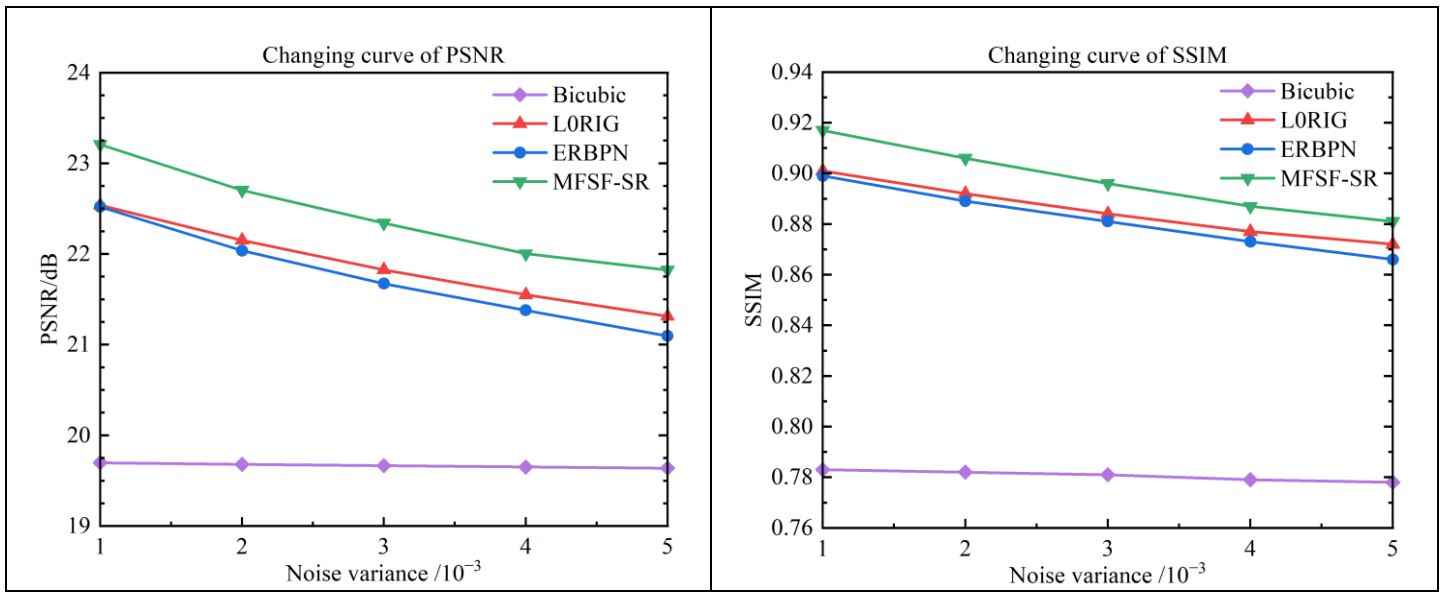

**Figure 9.** Quantitative comparison for reconstruction results under different Gaussian noise levels.

For these simulation experiments, Figure 10 shows the HR reconstruction results for the different methods at a scale factor of 4×. The green boxes show the zoomed regions to compare the performance of different methods. As the partial enlargement shows, the L0RIG method shows a better trade-off between removing noise and preserving the edges, but it is not able to recover the lost fine details. Undesired edge artifacts can be found in the results of the ERBPN method, which produces artificial edges in the flat surfaces and fails to suppress the noise in the details of the image. In Figure 10, the result of the proposed method shows a very good performance, with clear details and fewer ringing effects. Specifically, the distorted content, e.g., the stripes on the zebra, can be finely restored in the proposed two-step cascade model. Overall, the MFSF-SR cascade model performs favorably when compared to the baseline methods in this comparison experiment. It demonstrated that cascading L0RIG and ERBPN to enhance each individual baseline methods can substantially improve the final super-resolved image.

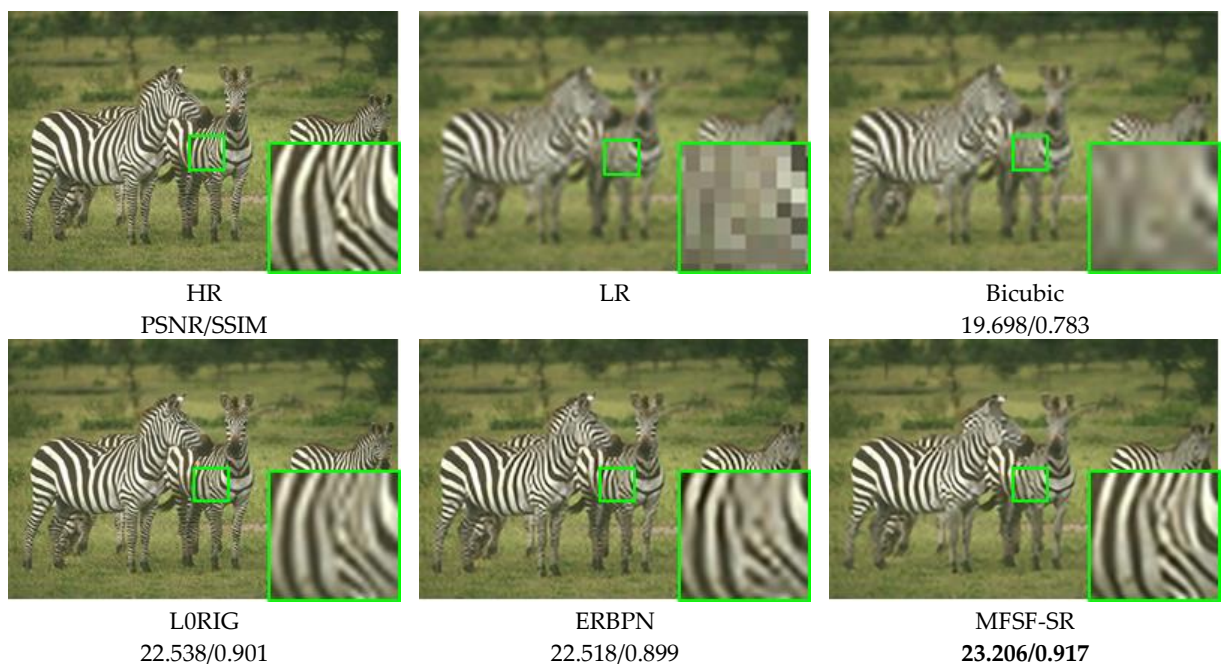

| HR | LR | Bicubic |
| PSNR/SSIM | | 19.698/0.783 |
| L0RIG | ERBPN | MFSF-SR |
| 22.538/0.901 | 22.518/0.899 | **23.206/0.917** |

**Figure 10.** Qualitative comparison of the zebra image under Gaussian noise with σ = 0.001.

In conclusion, with the qualitative and quantitative analysis, most of the results show that the cascade model creates more high-frequency information than the L0RIG and ERBPN methods. The MFSF-SR method works better in either noisy or noise-free case. It can reliably recover high-frequency details with higher consistency and contrast loss, while preserving strong edges and contours with few additional artifacts. The results were perceived as most informative and natural.

*3.3. Experiments on Real Data*

Besides the above experiments on synthetic test images, we also conducted experiments on real images to demonstrate the effectiveness of the proposed MFSF-SR cascade model. The real image grayscale sequences of Car and Eia are part of the Multi-Dimensional Signal Processing Research Group (MDSP) benchmark dataset [50], which is the most widely used dataset to test the performance of multi-frame super-resolution methods. In our experiment, 16 frames from these two image sequences were used as the low-resolution input image. The central frame in the sequence was set as the reference frame in this reconstruction.

Since no ground-truth HR image is available for the real sequence, we introduced no-reference image evaluation metrics the natural image quality evaluator (NIQE) [51] and the perception-based image quality evaluator (PIQE) [52] to further evaluate the quality of the real image SR results. Smaller values of NIQE and PIQE indicate better SR results. Figure 11 provides a visual comparison of the super-resolved results for the Car and Eia images with magnification factor 4. The red rectangles show zoomed regions of the restored images to compare the qualitative performance of the different methods. Experimental results on real image sequences show that our method yields a boosted performance in both objective metrics and visual quality. The MFSF-SR method achieves comparable or even better performance than the baseline methods in terms of quantitative evaluations. For a real-world image, the downsampling kernel is unknown and complicated; thus, performance of the non-blind SR methods are severely affected. Nevertheless, our method can produce visual pleasant images and effectively suppress the errors caused by noise, registration, and bad estimation of unknown PSF kernels.

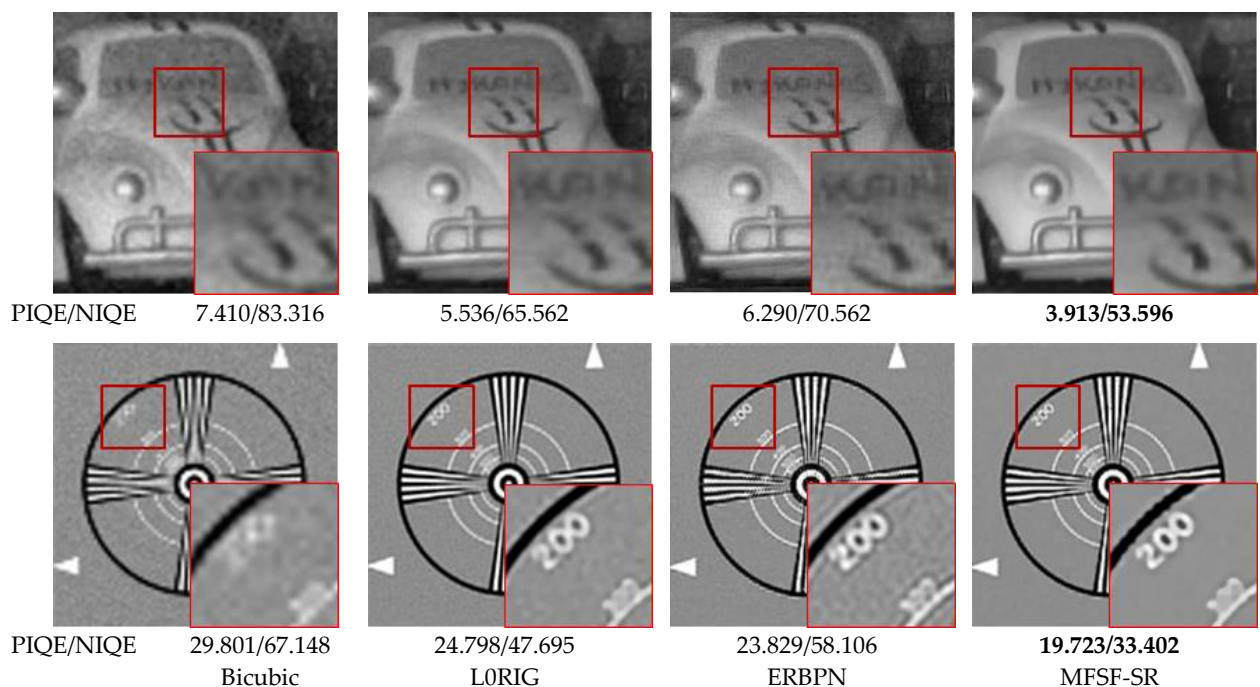

| PIQE/NIQE | 7.410/83.316 | 5.536/65.562 | 6.290/70.562 | **3.913/53.596** |
| PIQE/NIQE | 29.801/67.148 | 24.798/47.695 | 23.829/58.106 | **19.723/33.402** |
| | Bicubic | L0RIG | ERBPN | MFSF-SR |

**Figure 11.** Qualitative comparison of the Car and Eia images on 4× SR.

From the top line of Figure 11, we can observe that the experiment with the Car sequence can be considered as a challenging example because the LR Car images are severely degraded by blur and noise, with a complicated noise model. It was observed that the bicubic interpolation method is too blurry to be recognized, while the L0RIG and ERBPN algorithms can produce better visual effects than the bicubic interpolation method. Compared with the bicubic interpolation method, other methods are more efficient in improving spatial resolution due to the use of LR frame sequences or external prior knowledge in the reconstruction. With a L0-norm regularized constrain, L0RIG algorithm prefers a smooth result, but important edges and texture are also oversmoothed. As a contrast, the result of ERBPN suffers from visible ghosting artifacts and is seriously affected by the stair effects. As expected, the MFSF-SR algorithm has the best visual performance with clear edges and less influence of artifacts and can effectively remove noise in the smoothing area of the image. Meanwhile, as shown in the bottom line of Figure 11, the proposed method gives rise to the most visually pleasing results with both sharpness and naturalness. The L0RIG algorithm has a good noise suppression effect, but it over-smooths the image, resulting in the loss of edge information. In contrast, ERBPN produce result with sharp edges, but it lacks the ability to recover clean HR image because of the effect of artifacts. In summary, the proposed MFSF-SR cascade model is capable of generating clean and sharp HR images at a large scale factor without any hallucination of fine details. It consistently demonstrated the effectiveness and superiority in the thorough experiments conducted in this study.

## 4. Discussion

### 4.1. Effectiveness of the Two Different Cascade Models

The validity and reliability of the proposed MFSF-SR method was proven by the experiments described in Sections 3.2 and 3.3. To further investigate the effectiveness of the two different cascade models, we tested the two cascade models of MFSF-SR and SFMF-SR. The two kinds of cascade models combine the multi-frame-based L0RIG method with the single-frame-based ERBPN method in opposite order, compared with the two kinds of baseline methods with only one upsampling step at the 4× magnification factor. The SFMF-SR method reconstructed with ERBPN + L0RIG. Each LR image is independently enhanced using SFSR to obtain a higher-quality output. Then, the multi-frame-based L0RIG method is applied to the reconstructed images to obtain the final result for the reference image with a 2× scaling factor.

Table 3 shows the quantitative performance comparison in terms of the mean of PSNR and SSIM with the different cascade models on the Set5 [53] and Set14 [54]. On 4× enlargement, the cascade model, MFSF-SR, gains 0.339 dB and 0.364 dB more than SFMF-SR on the Set5 and Set14, respectively. It demonstrates that the cascade model by applying MFSR first and SFSR after outperforms the cascade model in the opposite order. Meanwhile, both of the two cascade models improve the quantitative performance compared to the two baseline methods of L0RIG and ERBPN. Figure 12 provides a visual comparison of simulation experiment results for the cameraman image with magnification factor 4. The images enclosed in red box show zoomed regions of the restored images to compare the qualitative performance of the different algorithms. As one can see, the cascade model of ERBPN + L0RIG tends to generate unexpected artifacts and seriously affected by the ringing effects. In fact, the MFSF-SR generates softer patterns containing more details and fewer blurred contours which subjectively closer to the ground truth. It produces superior results compared to the other cascade model in both objective and perceptual quality measurements. Additionally, the MFSF-SR approach also has significantly lower computational complexity than the SFMF-SR method that first applies SFSR to all the input LR images.

**Table 1.** Average PSNR/SSIM results for 4× SR on datasets Set5, Set14. Best and second best results are highlighted and underlined.

| Dataset | Metric | MFSR(L0RIG) | SFSR(ERBPN) | SFMF-SR(ERBPN + L0RIG) | MFSF-SR(L0RIG + ERBPN) |
|---|---|---|---|---|---|
| Set5 | PSNR | 31.962 | 32.653 | <u>33.075</u> | **33.413** |
| | SSIM | 0.891 | 0.899 | <u>0.910</u> | **0.917** |
| Set14 | PSNR | 28.354 | 29.037 | <u>29.294</u> | **29.658** |
| | SSIM | 0.779 | 0.791 | <u>0.821</u> | **0.828** |

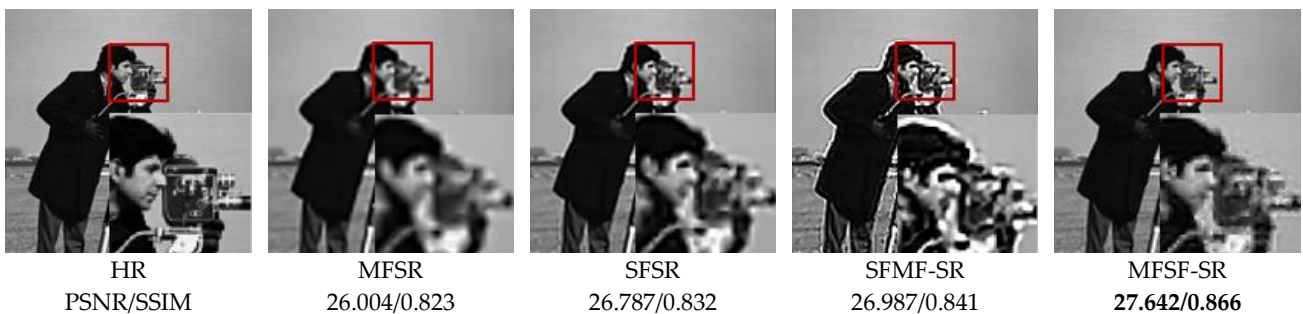

| HR | MFSR | SFSR | SFMF-SR | MFSF-SR |
|---|---|---|---|---|
| PSNR/SSIM | 26.004/0.823 | 26.787/0.832 | 26.987/0.841 | **27.642/0.866** |

**Figure 12.** Qualitative comparison of different methods for the cameraman image on 4× SR.

*4.2. Exploring the Robustness of Cascading Model*

In this section, we further discuss the generalization performance and limitations of the proposed cascade model. According to the above discussion, the MFSF-SR approach was selected as the proposed cascade model due to its better performance than the SFMF-SR approach. Another group of the state-of-the-art MFSR and SFSR methods were selected as the baseline methods, such as the MFSR based on the spatially weighted bilateral total variation regularization model (SWBTV) [19] and the SFSR method with the inaccurate kernel progressively correction (IKC) [55]. These two approaches were embedded into our MFSF-SR framework to verify the robustness of the cascade model.

There are four groups cascade methods with the combination of the four methods in a cascade manner of multiple first and single later. The simulation experiments include eight sets of comparative algorithms in addition to bicubic interpolation. The eight sets of comparison algorithms are single-frame- and multi-frame-based methods, as well as cascade methods: (1) the MFSR method of SWBTV [19] (denoted by M1); (2) the MFSR method of L0RIG (denoted by M2); (3) the SFSR method of IKC [55](denoted by S1); (4) the SFSR method of ERBPN (denoted by S2); (5) the MFSF-SR method of SWBTV + IKC (denoted by M1S1); (6) the MFSF-SR method of SWBTV + ERBPN (denoted by M1S2); (7) the MFSF-SR method of L0RIG + IKC (denoted by M2S1); (8) the MFSF-SR method of L0RIG + ERBPN (denoted by M2S2).

Table 4 shows the quantitative performance comparison in terms of the mean of PSNR and SSIM with the different methods on the three public benchmark datasets: Set5 [53], Set14 [54], and Urban100 [56]. The Set5 and Set14 datasets consist of natural scenes; the Urban100 set contains challenging urban scenes images with details in different frequency bands. We can draw some conclusion from the quantitative comparison. Firstly, all four groups of cascade methods are superior to their constituent single-frame and multi-frame super-resolution methods by a large margin. Therefore, it can be concluded that the proposed cascade model performs successfully and is robust to different SFSR and MFSR methods. Secondly, with the significant progress of image super-resolution achieved by deep learning, the deep learning-based SFSR approaches greatly improved the SR performance on synthetic LR images. Finally, as the initial input images of the learning based SFSR method, the results of the model-based MFSR are complex and var-ied. Nevertheless, IKC can handle complex degraded images through iterative correction of blur kernels, so it is more robust in the cascade model.

**Table 4.** Average PSNR/SSIM results for 4× SR on datasets Set5, Set14, and Urban100. Best and second best results are highlighted and underlined. M1, M2, S1, and S2 represent the SR methods of SWBTV [19], L0RIG, IKC [55], and ERBPN, respectively. M1S1, M1S2, M2S1, and M2S2 represent the cascade methods of SWBTV + IKC, SWBTV + ERBPN, L0RIG + IKC, and L0RIG + ERBPN, respectively.

| Dataset | Metric | Bicubic | MFSR | | SFSR | | MFSF-SR | | | |
|---|---|---|---|---|---|---|---|---|---|---|
| | | | M1 | M2 | S1 | S2 | M1S1 | M1S2 | M2S1 | M2S2 |
| Set5 | PSNR | 28.423 | 30.985 | 31.962 | 31.520 | 32.653 | 33.125 | 33.007 | **33.601** | <u>33.413</u> |
| | SSIM | 0.811 | 0.865 | 0.891 | 0.878 | 0.899 | 0.912 | 0.909 | **0.921** | <u>0.917</u> |
| Set14 | PSNR | 26.101 | 27.703 | 28.354 | 28.263 | 29.037 | 29.349 | 29.258 | **29.813** | <u>29.658</u> |
| | SSIM | 0.704 | 0.757 | 0.779 | 0.774 | 0.791 | 0.824 | 0.818 | **0.837** | <u>0.828</u> |
| Urban100 | PSNR | 23.152 | 24.614 | 25.683 | 25.334 | 26.086 | 26.858 | 26.672 | **27.163** | <u>27.072</u> |
| | SSIM | 0.659 | 0.729 | 0.773 | 0.759 | 0.803 | 0.815 | 0.812 | **0.830** | <u>0.827</u> |

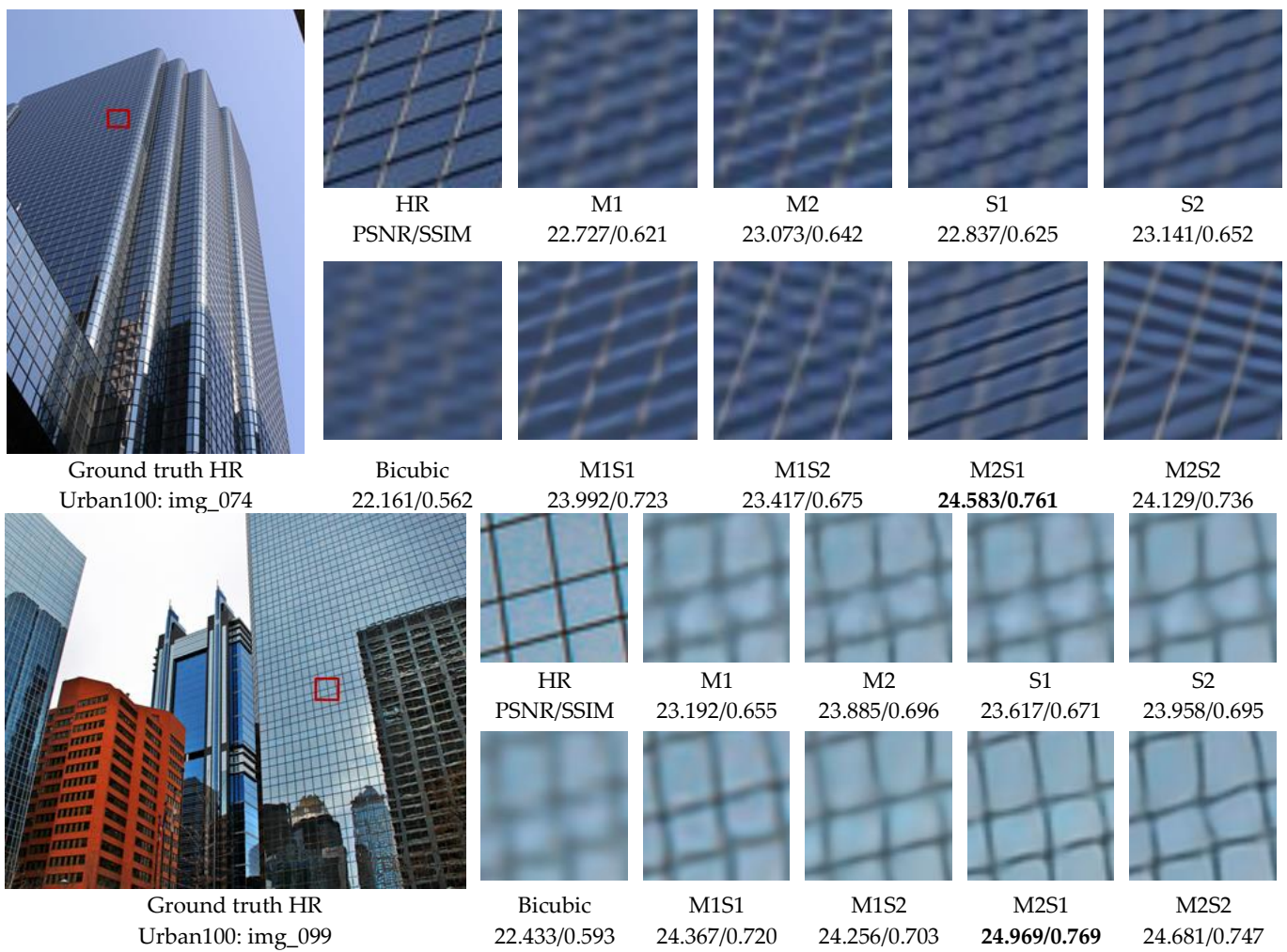

**Figure 13.** Visual comparison for 4× SR on the Urban100. M1, M2, S1, and S2 represent the SR methods of SWBTV [19], L0RIG, IKC [55], and ERBPN, respectively. M1S1, M1S2, M2S1, and M2S2 represent the cascade methods of SWBTV + IKC, SWBTV + ERBPN, L0RIG + IKC, and L0RIG + ERBPN, respectively.

In Figures 13, we show visual comparisons on Urban100 with a scale factor of 4× for the different comparative methods. Compare with the baseline methods, our cascade model more accurately reconstructs parallel straight lines, grid patterns such as windows. We obtain several observations from Figure 13. For image 'img_074' in Urban100, we can find that most baseline methods fail to recover edges and also suffer from blurring artifacts. Some of them even distort the horizontal lines and blur out the background. The results generated from IKC methods still contain noticeable artifacts caused by spatial aliasing. However, with an initialization reconstruction step by the model-based method of SWBTV or L0RIG, the cascade method SWBTV + IKC (M1S1) and L0RIG + IKC (M2S1) can effectively suppress such artifacts through progressive reconstruction. It significantly improves the performance of the resolved image with proper straight lines.

Similarly, in the second example, e.g., 'img_099' in Figure 13, the four baseline methods are unable to recover the rectangular shapes and blur out the boundaries representing the outlines of the windows. In contrast, the MFSF-SR cascade models show great abilities in producing accurate information from the LR image and removing the blur artifacts. Our method recovers the structures correctly with less distortion and more faithful to the ground-truth image. It was clearly demonstrated that the proposed cascade model can obtain a better tradeoff between recovering lost details and suppressing ringing artifacts. The abovementioned phenomena prove the advantages and robustness of the proposed cascade model on super-resolution reconstruction.

## 5. Conclusions

In this paper, we proposed a novel multi-frame super-resolution reconstruction concatenating the model-based MFSR method with the deep learning-based SFSR method. Our approach consists of an L0-norm constrained reconstruction scheme and an enhanced residual back-projection network in a concatenated fashion for image reconstruction. The proposed method first builds a MFSR method to obtain an initial result and apply SFSR method directly on the initial result. It takes both the sub-pixel shift information and external learned feature information into consideration, integrating the flexibility of the model-based method and the feature learning capacity of the deep learning-based method.

Extensive experiments on benchmark and real-world images illustrates that the proposed cascade model can significantly improve the performance of the super-resolution task. Superior results are produces compared to the other baseline methods in both qualitatively and quantitatively measurements. In addition, we have demonstrated that both the two kinds of cascade methods perform better than the baseline methods and the proposed cascade model can be robustly applied to different MFSR and SFSR methods. It means that potential future advances in MFSR and SFSR can be easily exploited to further improve the reconstructed image. In our future work, we will further study the coupling of the model-based MFSR and the deep learning-based SFSR methods in order to bring out their respective advantages.

**Author Contributions:** Conceptualization, J.S. and H.S.; methodology, J.S.; software, J.S.; investigation, J.L. and Q.Y.; writing—original draft preparation, J.S.; writing—review and editing, J.S. and Q.Y.; project administration, L.Z.; funding acquisition, H.S. All authors have read and agreed to the published version of the manuscript.

**Funding:** This work was supported in part by the National Natural Science Foundation of China under Grant 41971303 and in part by the Joint Project of Hubei Natural Science Foundation under Grant 2022CFD018.

**Conflicts of Interest:** The authors declare no conflicts of interest.

## References

1. Bai, H.; Pan, J. Self-Supervised Deep Blind Video Super-Resolution. *IEEE Trans. Pattern Anal. Mach. Intell.* **2024**.

2. Wen, W.; Ren, W.; Shi, Y.; Nie, Y.; Zhang, J.; Cao, X. Video Super-Resolution via a Spatio-Temporal Alignment Network. *IEEE Trans. Image Process.* **2022**, *31*, 1761–1773.
3. Gao, G.; Xu, Z.; Li, J.; Yang, J.; Zeng, T.; Qi, G.-J. CTCNet: A CNN-Transformer Cooperation Network for Face Image Super-Resolution. *IEEE Trans. Image Process.* **2023**, *32*, 1978–1991.
4. Lyu, Q.; Shan, H.; Steber, C.; Helis, C.; Whitlow, C.; Chan, M.; Wang, G. Multi-Contrast Super-Resolution MRI through a Progressive Network. *IEEE Trans. Med. Imaging* **2020**, *39*, 2738–2749.
5. Zhang, H.; Zhang, L.; Shen, H. A Super-Resolution Reconstruction Algorithm for Hyperspectral Images. *Signal Process* **2012**, *92*, 2082–2096.
6. Qiu, Z.; Shen, H.; Yue, L.; Zheng, G. Cross-Sensor Remote Sensing Imagery Super-Resolution via an Edge-Guided Attention-Based Network. *ISPRS J. Photogramm. Remote Sens.* **2023**, *199*, 226–241, doi:10.1016/j.isprsjprs.2023.04.016.
7. Xiao, Y.; Yuan, Q.; Jiang, K.; He, J.; Wang, Y.; Zhang, L. From Degrade to Upgrade: Learning a Self-Supervised Degradation Guided Adaptive Network for Blind Remote Sensing Image Super-Resolution. *Inf. Fusion* **2023**, *96*, 297–311, doi:10.1016/j.inffus.2023.03.021.
8. Chen, H.; He, X.; Qing, L.; Wu, Y.; Ren, C.; Sheriff, R.E.; Zhu, C. Real-World Single Image Super-Resolution: A Brief Review. *Inf. Fusion* **2022**, *79*, 124–145, doi:10.1016/j.inffus.2021.09.005.
9. Zhang, K.; Gao, X.; Tao, D.; Li, X. Single Image Super-Resolution With Non-Local Means and Steering Kernel Regression. *IEEE Trans. Image Process.* **2012**, *21*, 4544–4556, doi:10.1109/TIP.2012.2208977.
10. Gao, S.; Zhuang, X. Bayesian Image Super-Resolution with Deep Modeling of Image Statistics. *IEEE Trans. Pattern Anal. Mach. Intell.* **2022**, *45*, 1405–1423.
11. Timofte, R.; De Smet, V.; Van Gool, L. A+: Adjusted Anchored Neighborhood Regression for Fast Super-Resolution. In Proceedings of the Asian Conference on Computer Vision; Springer, 2014; pp. 111–126.
12. Dong, C.; Loy, C.C.; He, K.; Tang, X. Image Super-Resolution Using Deep Convolutional Networks. *IEEE Trans Pattern Anal Mach Intell* **2016**, *38*, 295–307.
13. Tai, Y.; Yang, J.; Liu, X. Image Super-Resolution via Deep Recursive Residual Network. In Proceedings of the Proceedings of the IEEE conference on computer vision and pattern recognition; 2017; pp. 3147–3155.
14. Zhang, Y.; Tian, Y.; Kong, Y.; Zhong, B.; Fu, Y. Residual Dense Network for Image Super-Resolution. In Proceedings of the Proceedings of the IEEE conference on computer vision and pattern recognition; 2018; pp. 2472–2481.
15. Dai, T.; Cai, J.; Zhang, Y.; Xia, S.-T.; Zhang, L. Second-Order Attention Network for Single Image Super-Resolution. In Proceedings of the 2019 IEEE/CVF Conference on Computer Vision and Pattern Recognition (CVPR); IEEE: Long Beach, CA, USA, June 2019; pp. 11057–11066.
16. Tsai, R.Y.; Huang, T.S. Multiframe Image Restoration and Registration. *Multiframe Image Restor. Regist.* **1984**, *1*, 317–339.
17. Laghrib, A.; Ben-loghfyry, A.; Hadri, A.; Hakim, A. A Nonconvex Fractional Order Variational Model for Multi-Frame Image Super-Resolution. *Signal Process. Image Commun.* **2018**, *67*, 1–11, doi:10.1016/j.image.2018.05.011.
18. Hakim, M.; Ghazdali, A.; Laghrib, A. A Multi-Frame Super-Resolution Based on New Variational Data Fidelity Term. *Appl. Math. Model.* **2020**, *87*, 446–467.
19. Liu, X.; Chen, L.; Wang, W.; Zhao, J. Robust Multi-Frame Super-Resolution Based on Spatially Weighted Half-Quadratic Estimation and Adaptive BTV Regularization. *IEEE Trans. Image Process.* **2018**, *27*, 4971–4986.
20. Laghrib, A.; Hadri, A.; Hakim, A.; Raghay, S. A New Multiframe Super-Resolution Based on Nonlinear Registration and a Spatially Weighted Regularization. *Inf. Sci.* **2019**, *493*, 34–56.
21. Khattab, M.M.; Zeki, A.M.; Alwan, A.A.; Badawy, A.S. Regularization-Based Multi-Frame Super-Resolution: A Systematic Review. *J. King Saud Univ.-Comput. Inf. Sci.* **2020**, *32*, 755–762.
22. Yue, L.; Shen, H.; Li, J.; Yuan, Q.; Zhang, H.; Zhang, L. Image Super-Resolution: The Techniques, Applications, and Future. *Signal Process* **2016**, *128*, 389–408.
23. Timofte, R.; Rothe, R.; Van Gool, L. Seven Ways to Improve Example-Based Single Image Super Resolution. In Proceedings of the Proceedings of the IEEE conference on computer vision and pattern recognition; 2016; pp. 1865–1873.
24. Ayas, S.; Ekinci, M. Single Image Super Resolution Using Dictionary Learning and Sparse Coding with Multi-Scale and Multi-Directional Gabor Feature Representation. *Inf. Sci.* **2020**, *512*, 1264–1278.
25. Xu, Y.; Guo, T.; Wang, C. A Remote Sensing Image Super-Resolution Reconstruction Model Combining Multiple Attention Mechanisms. *Sensors* **2024**, *24*, 4492, doi:10.3390/s24144492.
26. Xiao, Y.; Yuan, Q.; Zhang, Q.; Zhang, L. Deep Blind Super-Resolution for Satellite Video. *IEEE Trans. Geosci. Remote Sens.* **2023**, *61*, 1–16, doi:10.1109/TGRS.2023.3291822.
27. Kim, J.; Lee, J.K.; Lee, K.M. Accurate Image Super-Resolution Using Very Deep Convolutional Networks. In Proceedings of the Proceedings of the IEEE conference on computer vision and pattern recognition; 2016; pp. 1646–1654.
28. Kim, J.; Lee, J.K.; Lee, K.M. Deeply-Recursive Convolutional Network for Image Super-Resolution. In Proceedings of the Proceedings of the IEEE conference on computer vision and pattern recognition; 2016; pp. 1637–1645.
29. Zhang, Y.; Li, K.; Li, K.; Wang, L.; Zhong, B.; Fu, Y. Image Super-Resolution Using Very Deep Residual Channel Attention Networks. In Proceedings of the Proceedings of the European conference on computer vision (ECCV); 2018; pp. 286–301.
30. Chan, K.C.; Zhou, S.; Xu, X.; Loy, C.C. Basicvsr++: Improving Video Super-Resolution with Enhanced Propagation and Alignment. In Proceedings of the Proceedings of the IEEE/CVF conference on computer vision and pattern recognition; 2022; pp. 5972–5981.

31. Molini, A.B.; Valsesia, D.; Fracastoro, G.; Magli, E. Deepsum: Deep Neural Network for Super-Resolution of Unregistered Multitemporal Images. *IEEE Trans. Geosci. Remote Sens.* **2019**, *58*, 3644–3656.
32. An, T.; Zhang, X.; Huo, C.; Xue, B.; Wang, L.; Pan, C. TR-MISR: Multiimage Super-Resolution Based on Feature Fusion with Transformers. *IEEE J. Sel. Top. Appl. Earth Obs. Remote Sens.* **2022**, *15*, 1373–1388.
33. Shen, H.; Jiang, M.; Li, J.; Zhou, C.; Yuan, Q.; Zhang, L. Coupling Model- and Data-Driven Methods for Remote Sensing Image Restoration and Fusion: Improving Physical Interpretability. *IEEE Geosci. Remote Sens. Mag.* **2022**, *10*, 231–249, doi:10.1109/MGRS.2021.3135954.
34. Wu, J.; Yue, T.; Shen, Q.; Cao, X.; Ma, Z. Multiple-Image Super Resolution Using Both Reconstruction Optimization and Deep Neural Network. In Proceedings of the 2017 IEEE Global Conference on Signal and Information Processing (GlobalSIP); IEEE: Montreal, QC, November 2017; pp. 1175–1179.
35. Kawulok, M.; Benecki, P.; Piechaczek, S.; Hrynczenko, K.; Kostrzewa, D.; Nalepa, J. Deep Learning for Multiple-Image Super-Resolution. *IEEE Geosci. Remote Sens. Lett.* **2019**, *17*, 1062–1066.
36. Lai, W.-S.; Huang, J.-B.; Ahuja, N.; Yang, M.-H. Fast and Accurate Image Super-Resolution with Deep Laplacian Pyramid Networks. *IEEE Trans. Pattern Anal. Mach. Intell.* **2019**, *41*, 2599–2613, doi:10.1109/TPAMI.2018.2865304.
37. Molina, R.; Vega, M.; Mateos, J.; Katsaggelos, A.K. Variational Posterior Distribution Approximation in Bayesian Super Resolution Reconstruction of Multispectral Images. *Appl. Comput. Harmon. Anal.* **2008**, *24*, 251–267.
38. Pan, R.; Reeves, S.J. Efficient Huber-Markov Edge-Preserving Image Restoration. *IEEE Trans. Image Process.* **2006**, *15*, 3728–3735.
39. Pan, J.; Hu, Z.; Su, Z.; Yang, M.H. L0-Regularized Intensity and Gradient Prior for Deblurring Text Images and Beyond. *IEEE Trans Pattern Anal Mach Intell* **2017**, *39*, 342–355, doi:10.1109/TPAMI.2016.2551244.
40. Jiang H.; Sun J.; Xie C.; Lai S.; Shen H. Super-resolution image reconstruction method under joint constraints of external and internal gradient. *Comput. Eng.* **2022**, *48*, 220–227.
41. Xu, L.; Jia, J.; Matsushita, Y. Motion Detail Preserving Optical Flow Estimation. *IEEE Trans Pattern Anal Mach Intell* **2012**, *34*, 1744–1757.
42. Boyd, S.; Parikh, N.; Chu, E.; Peleato, B.; Eckstein, J. Distributed Optimization and Statistical Learning via the Alternating Direction Method of Multipliers. *Found Trends Mach Learn* **2011**, *3*, 1–122.
43. Haris, M.; Shakhnarovich, G.; Ukita, N. Deep Back-Projection Networks for Single Image Super-Resolution. *IEEE Trans. Pattern Anal. Mach. Intell.* **2021**, *43*, 4323–4337, doi:10.1109/TPAMI.2020.3002836.
44. Wang, Z.; Chen, J.; Hoi, S.C. Deep Learning for Image Super-Resolution: A Survey. *IEEE Trans. Pattern Anal. Mach. Intell.* **2020**, *43*, 3365–3387.
45. Yu, Y.; Li, X.; Liu, F. E-DBPN: Enhanced Deep Back-Projection Networks for Remote Sensing Scene Image Superresolution. *IEEE Trans. Geosci. Remote Sens.* **2020**, *58*, 5503–5515, doi:10.1109/TGRS.2020.2966669.
46. Timofte, R.; Agustsson, E.; Van Gool, L.; Yang, M.-H.; Zhang, L. Ntire 2017 Challenge on Single Image Super-Resolution: Methods and Results. In Proceedings of the Proceedings of the IEEE conference on computer vision and pattern recognition workshops; 2017; pp. 114–125.
47. Kingma, D.P.; Ba, J. Adam: A Method for Stochastic Optimization. *ArXiv Prepr. ArXiv14126980* **2014**.
48. Wang, Z.; Bovik, A.C.; Sheikh, H.R.; Simoncelli, E.P. Image Quality Assessment: From Error Visibility to Structural Similarity. *IEEE Trans. Image Process.* **2004**, *13*, 600–612.
49. Roth, S.; Black, M.J. Fields of Experts. *Int. J. Comput. Vis.* **2009**, *82*, 205–229, doi:10.1007/s11263-008-0197-6.
50. Milanfar, P.; Farsiu, S. *MDSP Super-Resolution and Demosaicing Datasets*; 2013; https://users.soe.ucsc.edu/~milanfar/software/sr-datasets.html.
51. Mittal, A.; Soundararajan, R.; Bovik, A.C. Making a "Completely Blind" Image Quality Analyzer. *IEEE Signal Process. Lett.* **2013**, *20*, 209–212, doi:10.1109/LSP.2012.2227726.
52. N, V.; D, P.; Bh, M.C.; Channappayya, S.S.; Medasani, S.S. Blind Image Quality Evaluation Using Perception Based Features. In Proceedings of the 2015 Twenty First National Conference on Communications (NCC); February 2015; pp. 1–6.
53. Bevilacqua, M.; Roumy, A.; Guillemot, C.; Morel, M.A. Low-Complexity Single-Image Super-Resolution Based on Nonnegative Neighbor Embedding. In Proceedings of the Procedings of the British Machine Vision Conference 2012; British Machine Vision Association: Surrey, 2012; p. 135.1-135.10.
54. Zeyde, R.; Elad, M.; Protter, M. On Single Image Scale-up Using Sparse-Representations. In Proceedings of the Curves and Surfaces: 7th International Conference, Avignon, France, June 24-30, 2010, Revised Selected Papers 7; Springer, 2012; pp. 711–730.
55. Gu, J.; Lu, H.; Zuo, W.; Dong, C. Blind Super-Resolution With Iterative Kernel Correction. In Proceedings of the IEEE/CVF conference on computer vision and pattern recognition; 2019; pp. 1604–1613.
56. Huang, J.-B.; Singh, A.; Ahuja, N. Single Image Super-Resolution from Transformed Self-Exemplars. In Proceedings of the Proceedings of the IEEE conference on computer vision and pattern recognition; 2015; pp. 5197–5206.